\documentstyle[aps,epsf,psfig]{revtex}

 \newcommand{\eins}{\mbox{$1 \hspace{-1.0mm}  {\bf l}$}}
 \newcommand{\be}{\begin{equation}}
 \newcommand{\ee}{\end{equation}}
 \newcommand{\bea}{\begin{eqnarray}}
 \newcommand{\eea}{\end{eqnarray}}
 \newcommand{\half}{\mbox{$\textstyle \frac{1}{2}$}}

 \newcommand{\ket}[1]{ | \, #1  \rangle}
 \newcommand{\bra}[1]{ \langle #1 \,  |}

 \newcommand{\braket}[2]{\left< #1 \right| #2 \rangle}

 \newcommand{\mytext}[1]{\mbox{ #1}}

\begin{document}
  \begin{center}
                 {\bf  Optimal cloning for two pairs of orthogonal
                          states }\newline
  
                 {Dagmar~Bru\ss $^1$ 
                  and Chiara~Macchiavello$^2$}\newline
                 {\em
                 $^1$Inst. f\"{u}r Theoret. Physik, Universit\"{a}t Hannover,
                 Appelstr. 2, D-30167 Hannover, Germany\\
                 $^2$Dipartimento di Fisica ``A. Volta'' and INFM-Unit\`a 
                 di Pavia,
                 Via Bassi 6, 27100 Pavia, Italy}
  \end{center}
                 \date{Received \today}
                  \begin{abstract}

We study the optimal cloning transformation for two pairs of orthogonal
states of two-dimensional quantum systems, and derive the corresponding
optimal fidelities.   

                 \end{abstract}
                PACS {03.67, 03.65} \newline

The possibility of cloning quantum states approximately  has attracted
much attention in the recent years. Limits for the efficiency of cloning
transformations have been derived in various cases.
For two-dimensional quantum systems the optimal fidelity 
has been reported for universal cloning \cite{buzek-hill,gima,bc,werner}, 
where the cloning transformation is optimised for the case that
all possible  
 states from the Bloch sphere are  treated in the same way;
  for phase covariant 
cloning
\cite{pcc}, where the fidelity is 
optimised for states lying on a great-circle of the Bloch
sphere, and for the case of just two non-orthogonal states \cite{oxibm}.

\begin{figure}[hbt]
\setlength{\unitlength}{1pt}
\begin{picture}(500,200)
\put(300,150){x}
\put(180,270){z}
\put(188,178){$\varphi$}
\put(240,230){$\vec m_1$}
\put(120,230){$\vec m_2$}
\put(120,70){$\vec m_3$}
\put(240,70){$\vec m_4$}
\epsfysize=10cm
\epsffile[72 230 540 560]{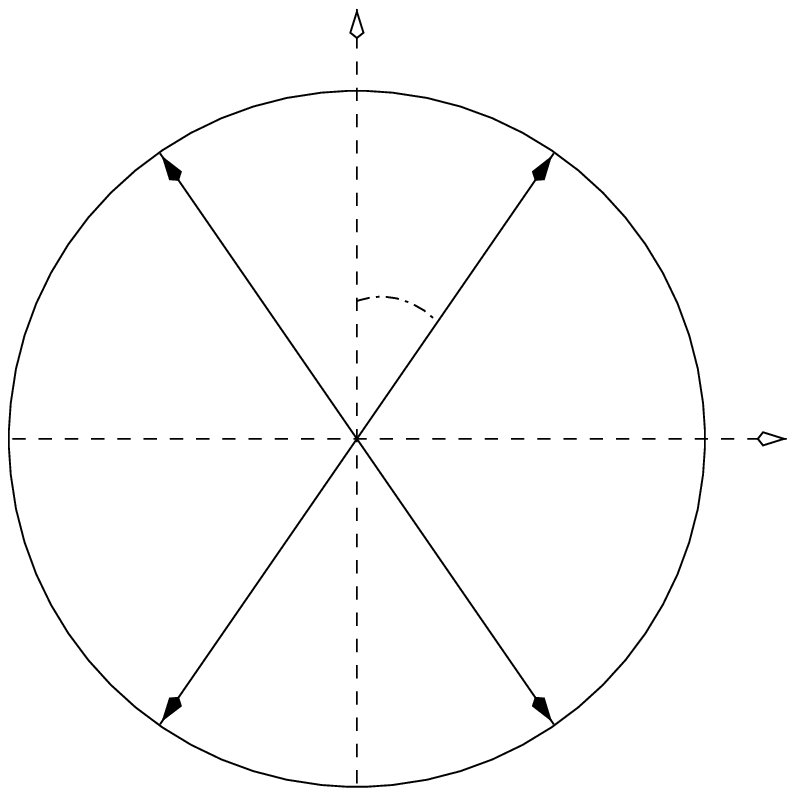}
\end{picture}
\vspace{-2cm}
\caption[]{\small Geometrical disposition of two pairs of orthogonal states.
  }
\label{pair}
\end{figure}

In this paper we study the optimal cloning for an ensemble of input states 
that consists of two pairs of orthogonal states for a two-dimensional quantum
system. These four states can be parametrized in the Bloch sphere 
representation with a single parameter in the following way.
The four Bloch vectors ${\vec m}_i$ for the states $\ket{\psi_i}$ with 
\be
\ket{\psi_i}\bra{\psi_i}
 = \half (\eins + {\vec m}_i \cdot {\vec \sigma}) \ \ \ \ \ \ i=1,...,4\;,
\ee
where $\eins$ is the identity operator and $\sigma_i$ with $i=x,y,z$ 
are the Pauli
matrices, are given by
\be
{\vec m}_1 = \left( \begin{array}{c} \sin{\varphi} \\
                                              0  \\
                        \cos{\varphi}    \end{array}\right) \ \ , \ \ 
{\vec m}_2 = 
          \left( \begin{array}{c} -\sin{\varphi} \\
                                              0  \\
                        \cos{\varphi}    \end{array}\right) \ \ , \ \
{\vec m}_3 = 
          \left( \begin{array}{c} -\sin{\varphi} \\
                                              0  \\
                        -\cos{\varphi}    \end{array}\right) \ \ , \ \
{\vec m}_4 = 
          \left( \begin{array}{c} \sin{\varphi} \\
                                              0  \\
                        -\cos{\varphi}    \end{array}\right) \ .
\label{ensemble}
\ee
In this representation the four vectors 
are lying in the $x,z$-plane, and each of them includes an
angle $\pm \varphi$ or $\pm (\pi -\varphi)$ with the $z$-axis,
see figure \ref{pair}. 

\vspace{2.5cm}

The two pairs of orthogonal states are given by 
$\{\ket{\psi_1},\ket{\psi_3}\}$ and $\{\ket{\psi_2},\ket{\psi_4}\}$.
\par
We could also parametrize  the states $\ket{\psi_i}$ with the real parameters
$\alpha$ and $\beta$ with $\alpha^2+\beta^2=1$:
\be
\ket{\psi_1} = \alpha \ket{0} +\beta \ket{1}\ \ , \ \
\ket{\psi_2} = \alpha \ket{0} -\beta \ket{1} \ \ ,\ \
\ket{\psi_3} = \beta \ket{0} -\alpha \ket{1} \ \ , \ \
\ket{\psi_4} =\beta \ket{0} +\alpha \ket{1} \ ,
\ee
where the relation between the parameters $\alpha$ and $\varphi$ 
is given by
\be
\alpha = \cos{\frac{\varphi}{2}} \ .
\ee
We study the case of $1\to 2$ cloning, namely two output copies are produced
from a single input. 
In this case
a cloning transformation is generally described  by a unitary operation 
acting on the input, a prescribed blank qubit,  and an auxiliary system,
initially in an arbitrary state $\ket{X}$.
In order to derive the optimal cloning transformation, it is sufficient 
to define its action on the basis states of the input, namely
\bea
U\, \ket{0}\ket{0}\ket{X} & = & a\ket{00}\ket{A}+b(\ket{01}
                      +\ket{10})\ket{B}+ c\ket{11}\ket{C},
                  \nonumber \\
                 U\, \ket{1}\ket{0}\ket{X} & = & \tilde a\ket{11}
                          \ket{\tilde A}+\tilde b(\ket{10}
                      +\ket{01})\ket{\tilde B}
                      + \tilde c\ket{00}\ket{\tilde C},
                   \label{eq:10}
\eea
where the coefficients $a,b,c$ 
can be taken real and positive
by including possible phases into the ancilla states. 
The above form for the cloning transformation guarantees that the two
output copies are described by the same reduced density operator.
We study cloning transformations that lead to the same efficiency for
the four states $\ket{\psi_i}$. Since the four states are transformed into one
another by renaming the basis states, i.e.
$\ket{0}\leftrightarrow\ket{1}$, the cloning transformation will be invariant
under the exchange of $\ket{0}$ and $\ket{1}$. This condition 
leads to $a={\tilde a}, b={\tilde b}, c={\tilde c}$.
Moreover, unitarity of the cloning transformation $U$ dictates the condition 
\be
a^2+2b^2+c^2=1 \ .
\label{uni}
\ee
We describe the efficiency of the cloning transformation in terms of the 
fidelity $F$ of each output copy with respect to the input state, namely
\be
F=\bra{\psi_i}\rho_i\ket{\psi_i}\;,
\ee
where $\rho_i={\mbox Tr}[U\ket{\psi_i}\bra{\psi_i}U^\dagger]$ and the trace
is performed over the auxiliary system and one of the output copies.
With our symmetric way to parametrize the states we can easily derive 
the fidelity for the four input states,
as we just have to calculate the fidelity once and can then use
symmetry arguments in order to find the explicit form of the other three
cases, e.g. we can replace
$\beta$ by $-\beta$ to go from the fidelity for $\ket{\psi_1}$ to the
fidelity for $\ket{\psi_2}$.
 \par
As mentioned above, we require the four fidelities to be equal. This
condition leads to 
\be
F=a^2(\alpha^4+\beta^4)+2c^2\alpha^2\beta^2+b^2
+\alpha^2\beta^2[ab\cdot 2 \mytext{Re}(\braket{A}{\tilde B}+
  \braket{B}{\tilde A})+
  bc\cdot 2 \mytext{Re}(\braket{B}{\tilde C}+
  \braket{C}{\tilde B})] \ .
\label{fideli}
\ee
Independently  of the coefficients $a,b,c$ the fidelity
will be maximal for the following choice of scalar products between
the auxiliary states:
\bea
   \braket{A}{\tilde B}&=&1=
  \braket{B}{\tilde A}     \ ,\nonumber \\
   \braket{B}{\tilde C}&=&1=
  \braket{C}{\tilde B}     \ ,
\eea
which can be reached with a two-dimensional ancilla and, e.g., the 
choice
\bea
\ket{A}&=&\ket{0} \ \ , \ \ \ket{B}=\ket{1}\ \ , \ \ \ket{C}=\ket{0}\ ,
\nonumber \\
\ket{\tilde A}&=&\ket{1} \ \ , \ \ \ket{\tilde B}=\ket{0}\ \ , \ \ 
\ket{\tilde C}=\ket{1} \ .
\label{ancilla}
\eea
Inserting this into equation (\ref{fideli}) we arrive at
\be
F=\half + \half(a^2-c^2)\cos^2{\varphi}  + b(a+c)\sin^2{\varphi}  \ .
\label{fidelity}
\ee
The optimal cloning tranformation corresponds to the maximum value of the
fidelity (\ref{fidelity}), together with the constraint (\ref{uni})
due to unitarity.
\par
Using the method of Lagrange multipliers we thus have to solve the
system of equations
\bea
 a \cos^2{\varphi} + b\sin^2{\varphi}  & = & 2 a \lambda \ ,\nonumber \\
(a+c)\sin^2{\varphi}  & = & 4 b \lambda \ ,\nonumber \\
- c \cos^2{\varphi} + b\sin^2{\varphi}  & = & 2 c \lambda \ ,\nonumber \\
a^2+2b^2+c^2 & = & 1\ \ ,
\eea
where $\lambda$ is the Lagrange multiplier.
The solution for the coefficients $a,b$ and $c$ turns out to be
\bea
a & = & \half(1 + \cos^2{\varphi} \sqrt{\frac{1}{\sin^4{\varphi}+\cos^4{\varphi}}}\, ) \ ,\nonumber \\
b & = & \half \sin^2{\varphi}  \sqrt{\frac{1}{\sin^4{\varphi}+\cos^4{\varphi}}} \ , \nonumber \\
c & = & \half(1 - \cos^2{\varphi} \sqrt{\frac{1}{\sin^4{\varphi}+\cos^4{\varphi}}}\, ) \ \ .
\label{coeff}
\eea 
  Inserting this into equation (\ref{fidelity}) leads to the optimum fidelity 
\bea
F^{opt} = \half(1+\sqrt{\sin^4\varphi+\cos^4\varphi}) \ .
\eea      
The explicit form of the resulting optimal cloning transformation is found
immediately by inserting equations (\ref{coeff}) and (\ref{ancilla}) 
into equation (\ref{eq:10}).

\begin{figure}[hbt]
\setlength{\unitlength}{1pt}
\begin{picture}(500,200)
\put(380,8){$\varphi$}
\put(130,125){$F^{opt}$}
\epsfysize=10cm
\centerline{\psfig{file=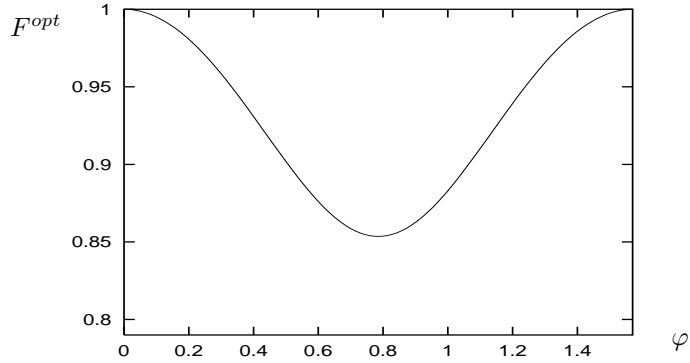,height=5cm,width=8cm}}
\vspace{-0.2cm}
\end{picture}
\vspace{0.5cm}
\caption[]{\small Optimal fidelity 
 for cloning  two pairs of orthogonal states, as a function of $\varphi$.
  }
\label{fig}
\end{figure}

In figure \ref{fig} we plot $F^{opt}$ as a function of the angle
$\varphi$. 
The figure demonstrates that the cloning task is performed in the worst 
way for the
two pairs being maximally spread, i.e. in the  case 
$\varphi=\pi/4$. This is the well-known setting for quantum cryptography
in the BB84 scheme \cite{crypto}. 
Notice that  value for the optimal fidelity of cloning in the
BB84  scheme, derived in \cite{pcc}, is recovered here. 
As the ability to make approximate clones of a state is related to the
security of a cryptographic protocol, our calculations
indicate that the BB84 scheme is the most favourable setting
when using four states for cryptography.


\vspace{-2cm}

\begin{figure}[hbt]
\setlength{\unitlength}{1pt}
\begin{picture}(500,200)
\put(380,8){$\varphi$}
\put(230,115){$\eta_z$}
\put(230,65){$\eta_x$}
\epsfysize=10cm
\centerline{\psfig{file=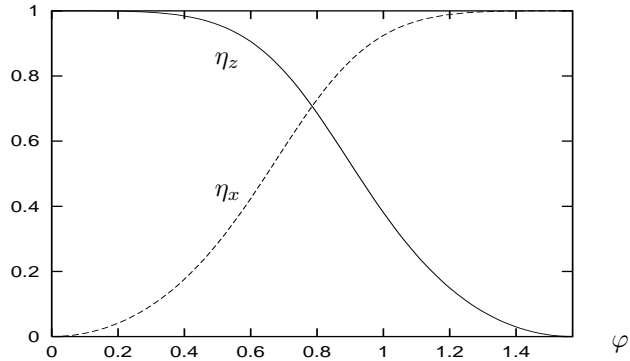,height=5cm,width=8cm}}
\vspace{-0.2cm}
\end{picture}
\vspace{0.5cm}
\caption[]{\small Shrinking factors  $\eta_x$ and $\eta_z$
 for cloning  two pairs of orthogonal states, as a function of $\varphi$.
  }
\label{shrink}
\end{figure}

We point out the following
 geometrical description of the cloning transformation.
For states with a Bloch vector lying on the $x-z$ plane of the Bloch sphere,
 namely states given by the density operator
$\rho=\frac{1}{2}(\eins+m_x\sigma_x+m_z\sigma_z)$, 
we can describe the cloning transformation (\ref{eq:10}) 
in terms of two shrinking
factors $\eta_x$ for the $x$-component of the Bloch vector,
and $\eta_z$ for its $z$-component, such that
 the output state of each copy takes
the form $\rho_{out}=
\frac{1}{2}(\eins+\eta_x m_x\sigma_x+\eta_z m_z\sigma_z)$. 
The explicit expression for the two shrinking factors 
with our choice of ancillas (\ref{ancilla}) is given by
\bea
\eta_x &=& 2b(a+c) \ , \nonumber \\ 
\eta_z &=& a^2-c^2 \ .
\eea 
In the case of the optimal transformation, according to equation 
(\ref{coeff}), the shrinking factors
depend only on the value of $\varphi$:
\bea
\eta_x &=& \sin^2{\varphi}\sqrt{\frac{1}{\sin^4{\varphi}+\cos^4{\varphi}}}
         \ , \nonumber \\ 
\eta_z &=& \cos^2{\varphi}\sqrt{\frac{1}{\sin^4{\varphi}+\cos^4{\varphi}}} \ .
\eea 
 These shrinking factors are reported in figure 
\ref{shrink}. 
Notice that they become equal  for $\varphi=\pi/4$, namely 
 $\eta_x(\pi/4)=\eta_z(\pi/4)=1/\sqrt{2}$, and according to the symmetry
of the input ensemble (\ref{ensemble}) that 
we used to perform the optimisation they are
related as $\eta_x(\varphi)=\eta_z(\pi/2-\varphi)$. Furthermore,
the identity $\eta_x^2+\eta_z^2=1$ holds.

\vspace{-2cm}
\par
In summary, we have found the best cloning transformation and the
corresponding optimal fidelity for cloning two pairs of orthogonal
states, and mentioned the implications for choosing such an ensemble
of states for quantum cryptography.
A further application of our results could be the connection to
state estimation for this set of inputs, as optimal quantum cloning can be
part of an optimal state estimation process.
    
This work was supported in part by the European Union project EQUIP
(contract IST-1999-11053) and by Ministero dell'Universit\`{a} e della 
Ricerca
Scientifica e Tecnologica under the project ``Quantum information
transmission and processing: quantum teleportation and error correction''.
DB acknowledges support from the ESF Programme QIT
and DFG-Schwerpunkt QIV.


\begin{thebibliography}{99}

\bibitem{buzek-hill} V.~Bu\v{z}ek and M.~Hillery, Phys. Rev. A {\bf 54}, 
1844 (1996).
\bibitem{gima} N.~Gisin and S.~Massar, Phys.~Rev.~Lett. {\bf 79}, 2153
                   (1997).
\bibitem{bc} D.~Bru\ss , A.~Ekert, C.~Macchiavello, 
         Phys. Rev. Lett.  {\bf 81}, 2598 (1998).
\bibitem{werner} R. Werner, Phys.~Rev. A{\bf 58}, 1827 (1998).
\bibitem{pcc} D.~Bru\ss, M. Cinchetti, G.M. D'Ariano and C.~Macchiavello,
Phys. Rev. A {\bf 62}, 12302 (2000).
\bibitem{oxibm} D.~Bru\ss, D.~DiVincenzo, A.~Ekert, C.~Fuchs,
             C.~Macchiavello
               and J.~Smolin, Phys.~Rev. A {\bf 57}, 2368 (1998).
\bibitem{crypto}C.~H.~Bennett and G.~Brassard, 
{\em Proc. IEEE
Int. Conf. on Computers, Systems, and Signal Processing, Bangalore,
India} (IEEE, New York, 1984),  175.
\end{thebibliography}
\end{document}